\documentclass[sigconf,nonacm]{acmart}
\renewcommand\footnotetextcopyrightpermission[1]{} 
\usepackage{natbib}
\usepackage{graphicx}
\graphicspath{ {./images/} }
\usepackage[font={footnotesize},skip=0.2pt]{caption}
\usepackage{setspace}
\usepackage{subcaption}
\usepackage{siunitx}
\newcolumntype{d}{S[
    input-open-uncertainty=,
    input-close-uncertainty=,
    parse-numbers = false,
    table-align-text-pre=false,
    table-align-text-post=false
 ]}
\AtBeginDocument{%
  \providecommand\BibTeX{{%
    \normalfont B\kern-0.5em{\scshape i\kern-0.25em b}\kern-0.8em\TeX}}}

\newcommand{\blfootnote}[1]{\begingroup\renewcommand{\thefootnote}{}\footnote{#1}\addtocounter{footnote}{-1}\endgroup}
\begin{document}

\title{Hierarchical Clustering As a Novel Solution to the Notorious Multicollinearity Problem in Observational Causal Inference}

\author{Yufei Wu}
\affiliation{%
  \institution{Airbnb, Inc.}
  \city{San Francisco}
  \state{California}
  \country{USA}
}
\email{yufei.wu@airbnb.com}

\author{Zhiying Gu}
\affiliation{%
  \institution{Airbnb, Inc.}
  \city{San Francisco}
  \state{California}
  \country{USA}
}
\email{zhiying.gu@airbnb.com}

\author{Alex Deng}
\affiliation{%
  \institution{Airbnb, Inc.}
  \city{San Francisco}
  \state{California}
  \country{USA}
}
\email{alex.deng@airbnb.com}

\author{Jacob Zhu}
\affiliation{%
  \institution{Airbnb, Inc.}
  \city{San Francisco}
  \state{California}
  \country{USA}
}
\email{jacob.zhu@airbnb.com}

\author{Linsha Chen}
\affiliation{%
  \institution{Airbnb, Inc.}
  \city{San Francisco}
  \state{California}
  \country{USA}
}
\email{linsha.chen@airbnb.com}

\renewcommand{\shortauthors}{Yufei Wu, Zhiying Gu, et al.}

\begin{abstract}
  Multicollinearity is a long lasting challenge in observational causal inference, especially under regressional settings — highly correlated independent variables make it difficult to isolate their individual impacts on outcomes of interest. While common solutions such as shrinkage estimators, principal component regressions, and partial linear regression are helpful in prediction problems, a crucial limitation hinders their applicability to causal inference problems — they cannot provide the original causal relationships. To fill the gap, we present an innovative and intuitive solution, by employing hierarchical clustering to aggregate data in a way that effectively alleviates collinearity. This method is generally applicable to causal problems featuring multicollinearity. We use a marketing application to demonstrate how and why it works.

  Expenditures on different advertising channels often exhibit correlations, making it exceedingly difficult to separately measure their impact. Many previous studies proposed to leverage granular cross-sectional data for better identification but, to our knowledge, none explicitly addressed multicollinearity, which undermines causal identification even with granular data. We propose to hierarchically cluster geographic units based on marketing spend correlation to reduce collinearity, and to implement a Bayesian Marketing Mix Model with cluster-level data. Such clustering happens in two steps — we first normalize and demean geo-level data to establish a common scale and to eliminate the common trends; we then calculate pairwise distance to summarize marketing spend correlation between geos and cluster the ones with moderate to strong correlation. Both descriptive evidence and regression analysis affirm that such hierarchical clustering effectively mitigates collinearity and facilitates the separate identification of the impact of different marketing channels.

\end{abstract}

\keywords{Marketing Mix Model, Hierarchical Bayesian, Hierarchical Clustering, Multicollinearity}

\maketitle
\blfootnote{\textcopyright\ 2023 Airbnb, Inc. All rights reserved.}

\section{Introduction} \label{Intro}
Everyday business inquiries frequently revolve around causal inference, specifically seeking to understand the impact of particular business decisions. To address this, three common approaches are typically employed: A/B testing, quasi-experimentation, and observational causal inference methods. While A/B testing and quasi-experimentation are often preferred due to their ability to provide exogenous variation for identification, their implementation can be prohibitively costly or subject to biases resulting from business or technical constraints. Observational causal inference methods, such as matching methods, synthetic control, and double machine learning are designed to mitigate biases, but are not applicable to some questions we aim to address given the data properties. Furthermore, the aforementioned approaches are more effective in measuring the causal impact of single interventions rather than attributing causal impact holistically across multiple interconnected factors that may contribute to the final outcome. In such scenarios, regression approaches provide a more suitable alternative. Regression methods only necessitate aggregated panel data to concurrently identify the causal impact of multiple factors. However, two challenges undermine our ability to confidently affirm that the estimated parameters from regression methods represent the true causal impact. These challenges are the existence of confounding factors and multicollinearity among covariates. While common solutions such as shrinkage estimators, principal component regressions, and partial linear regression are helpful in prediction problems, a crucial limitation hinders their applicability to causal inference problems — they cannot provide the original causal relationships.

This paper introduces a novel approach that specifically addresses the second challenge, namely multicollinearity. To illustrate the practical application and effectivenss of this approach, we demonstrate its implementation in a marketing measurement context (Marketing Mix Marketing) at Airbnb. We also conclude this paper with a discussion on the broad applicability of this approach.

In marketing, one question of paramount importance is to causally attribute sales to spend across channels - such as Google Search, YouTube, Display, etc. However, advertisers often allocate their expenditures across ad channels in a correlated manner, particularly during peak seasons. When attempting to estimate a regression model, highly correlated variables result in larger estimate variances and imprecise attribution of channel contributions to sales. It is not uncommon to observe regression coefficients switching signs when highly correlated inputs are introduced, consequently undermining the confidence of business stakeholders in the model results.

In this marketing application, we have access to panel data consisting of ad impressions categorized by channel and geographic location (Designated Market Area, or DMA) over a specific time period. When we analyze the data by pooling all geographic locations together, we observe a high level of cross-channel correlation. However, it is worth noting that certain geographic locations exhibit higher cross-channel correlations compared to others. To address the issue of multicollinearity, we propose a novel approach that leverages the variations in correlation patterns across different geographic locations. The objective is to restructure the data in a way that significantly reduces the multicollinearity problem. Our proposed method involves utilizing hierarchical clustering to group geographic locations based on their correlation patterns. The key aspect of this approach lies in defining the distance metric used in the clustering algorithm. 

In our methodology, we define the distance between two DMAs as the sum of channel-specific distances. Each channel-specific distance measures the similarity in the cross-channel correlation between the two geographic locations. By incorporating this distance metric into the hierarchical clustering process, we can effectively group the DMAs in a manner that minimizes multicollinearity across channels. This innovative approach allows us to transform the data structure, mitigating the challenges posed by multicollinearity and providing a more robust foundation for further analysis. By adopting this methodology, we can improve our understanding of the causal relationships between channels and accurately attribute their impact on sales or other relevant outcomes. We will demonstrate this improvement with both data descriptive evidence and regression results. 

The remainder of this paper is organized as follows. In section 2, we describe the Marketing Mix Modelling (MMM) problem formulation and the related work. In section 3, we present the data properties that motivate our approach to reduce multicollinearity. In section 4, we introduce Hierarchical Clustering and the distance metric designed specifically to address the multicollinearity problem. In section 5, we show how this novel method improves results, and in section 6, we briefly discuss other applications that can utilize this methodology.  

\section{Bayesian Structural Model Formulation} \label{BayesianModel}
This paper focuses on providing an innovative and intuitive solution to the notorious multicollinearity problem. To demonstrate the effectiveness of their approach, we apply it to a specific application called Bayesian Marketing Mix Modeling (MMM), built upon \citet{jin2017bayesian}. MMM is a widely applied method in the industry for estimating the performance of various marketing channels in a holistic manner. It takes into account factors such as seasonality, trend (representing organic demand), and mix of different marketing channels when forecasting sales. \footnote{While experimentation can be used to measure the performance of some channels, it is not always feasible due to practical constraints.}

\subsection{Marketing Mix Model Setup}
We model sales as a non-linear function of seasonality, and advertisement impressions of each channel with a Bayesian Model. Let $g$ denote a DMA and $t = 1, . . . , T$ denote time (we use weekly data). 
\[y_{g,t} = \mu t^\lambda + seasonality_{t} + \alpha Z_{g,t} + \sum_{k=1}^K \beta_k AdStock(x_{k, g,t})\ + \epsilon_{g,t}\]

There are $K$ media channels, and $G$ DMAs. $x_{g,t}$ is the impression of channel $k$ at week $t$. Let $y_{g,t}$ be the response variable at week $t$, which could be sales or log transformed sales. We include $\mu t^k$, $seasonality_{g,t}$, and contemporaneous correlation with covariates $Z_{g,t}$ to capture the evolution of organic sales over time. $AdStock(x_{g,t})$ is the transformed impressions that captures: (1) diminishing return; (2) lag of the effect; (3) carryover effect of the impressions. \citet{ng2021bayesian} uses a different formulation which only estimates the saturation effect. In this paper, we adapt Google's proposed shape formulation of marketing effects that is more flexible.\cite{jin2017bayesian} The $AdStock$ function can be defined as:

\[AdStock_{k,g} = \bigg(\frac{\sum_{l=0}^{L}\tau_{k}^{(l-\theta_{k})^2} x_{t-l,m}}{\sum_{l=0}^{L}\tau_{k}^{(l-\theta_{k})^2}}\bigg )^\rho \]

$\rho \in (0, 1]$ captures (potentially diminishing) returns to scale; $\tau \in (0, 1)$ governs the carryover rate over time; and $\theta$ indicates the lagged peak effect.

\subsection{Account for Confounding factors}
 
While it is not the main focus of this paper, we take into account confounding factors when modeling trend and seasonality in order to properly capture organic demand. When modeling trend and seasonality, it is crucial to strike the right balance between flexibility and strictness -- excessive flexibility may lead to overfitting, whereas overly rigid parametric formulations can result in a poor fit of the model. In this marketing use case, we can easily overfit a model that performs poorly out of sample because we have high dimensional parameters space - we have to estimate 4 parameters per channel. Keeping this tradeoff in mind, in addition to including exponential trend and sinusoidal seasonality following \citet{jin2017bayesian}, we also include as an additional covariate an index of Google Search query volume for travel and accommodation brands excluding Airbnb ($Z_{g,t}$ in the above equation), to capture confounding factors that affect organic demand contemporaneously.

\section{Data Properties} \label{Data}
\subsection{Pre-Process Data}
We take two important steps to pre-process data in preparation for descriptive analysis and modeling. First, we normalize bookings, channel impressions, and the covariate to establish a common scale across DMAs of different sizes. This will make it easier (A) to interpret the impact of a certain level of marketing activity; and (B) to model the common trend and seasonality later. Second, we decompose channel impressions into the common trend and seasonality and residual variation across DMAs, so we can focus on correlation in the residual variation next.  

\subsection{Characteristics of DMA-Level Data}
\begin{figure}[h]
  \caption{Correlation of Residual Channel Impressions}
  \centering
  \includegraphics[width=0.8\linewidth]{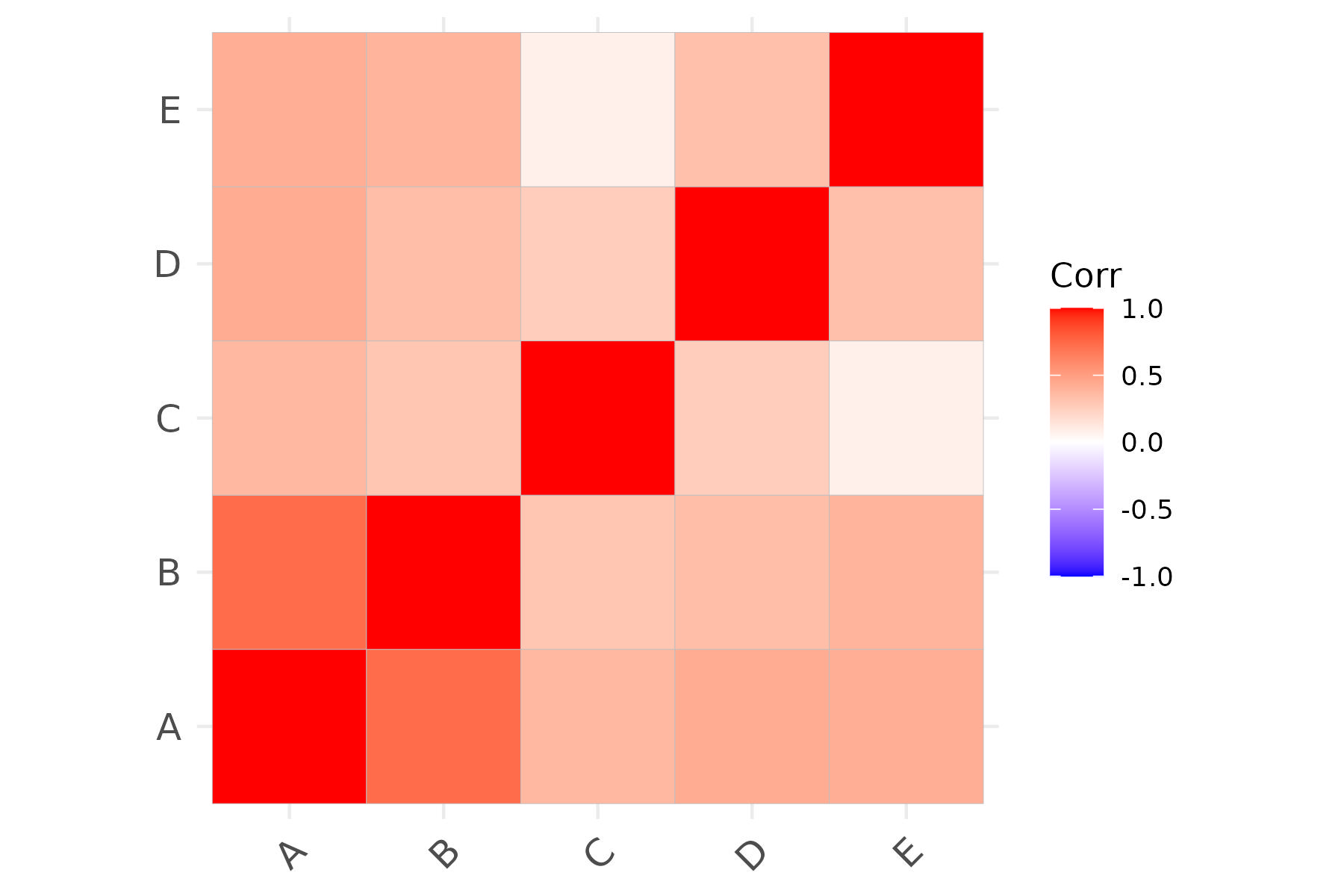}
  \label{fig:corr}
\end{figure}

There is a decent level of correlation between marketing channels and DMAs, even after eliminating the common trend and seasonality across DMAs for each channel. As Figure \ref{fig:corr} shows, the variation in residual impressions is moderately to strongly correlated across the five channels.\footnote{We anonymize the channels as A, B, C, D, and E.} 

\begin{figure}
  \centering
  \caption{Correlation of Residual Impressions Across DMAs}
    \begin{subfigure}[b]{0.5\textwidth}
        \centering
        \caption{Correlation Across DMAs in the Xth Ventile of Size: By Channel}
        \includegraphics[width=\linewidth]{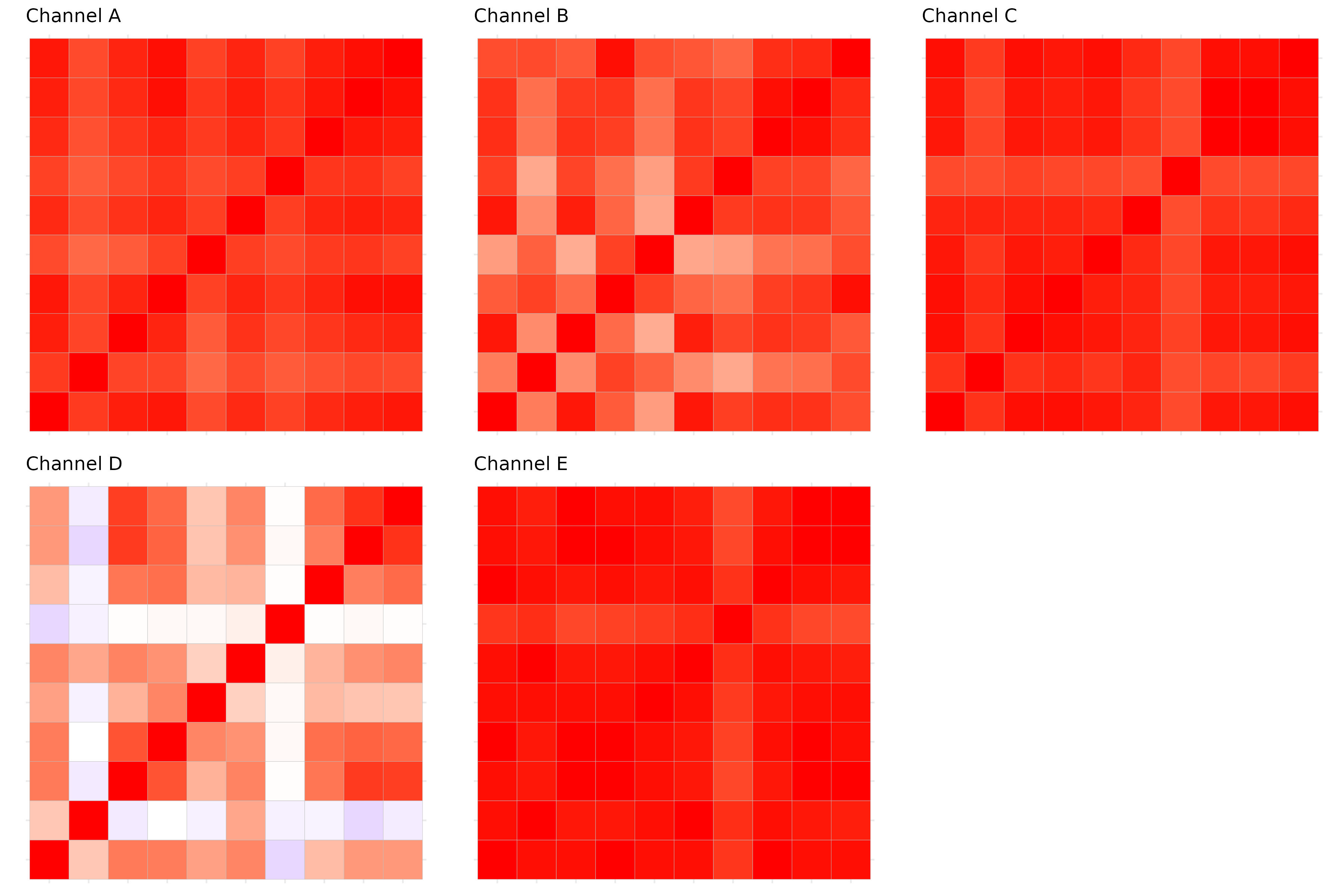}
    \end{subfigure}
    \hfill
    \begin{subfigure}[b]{0.5\textwidth}
        \centering
        \caption{Correlation Across DMAs in the Yth Ventile of Size: By Channel}
        \includegraphics[width=\linewidth]{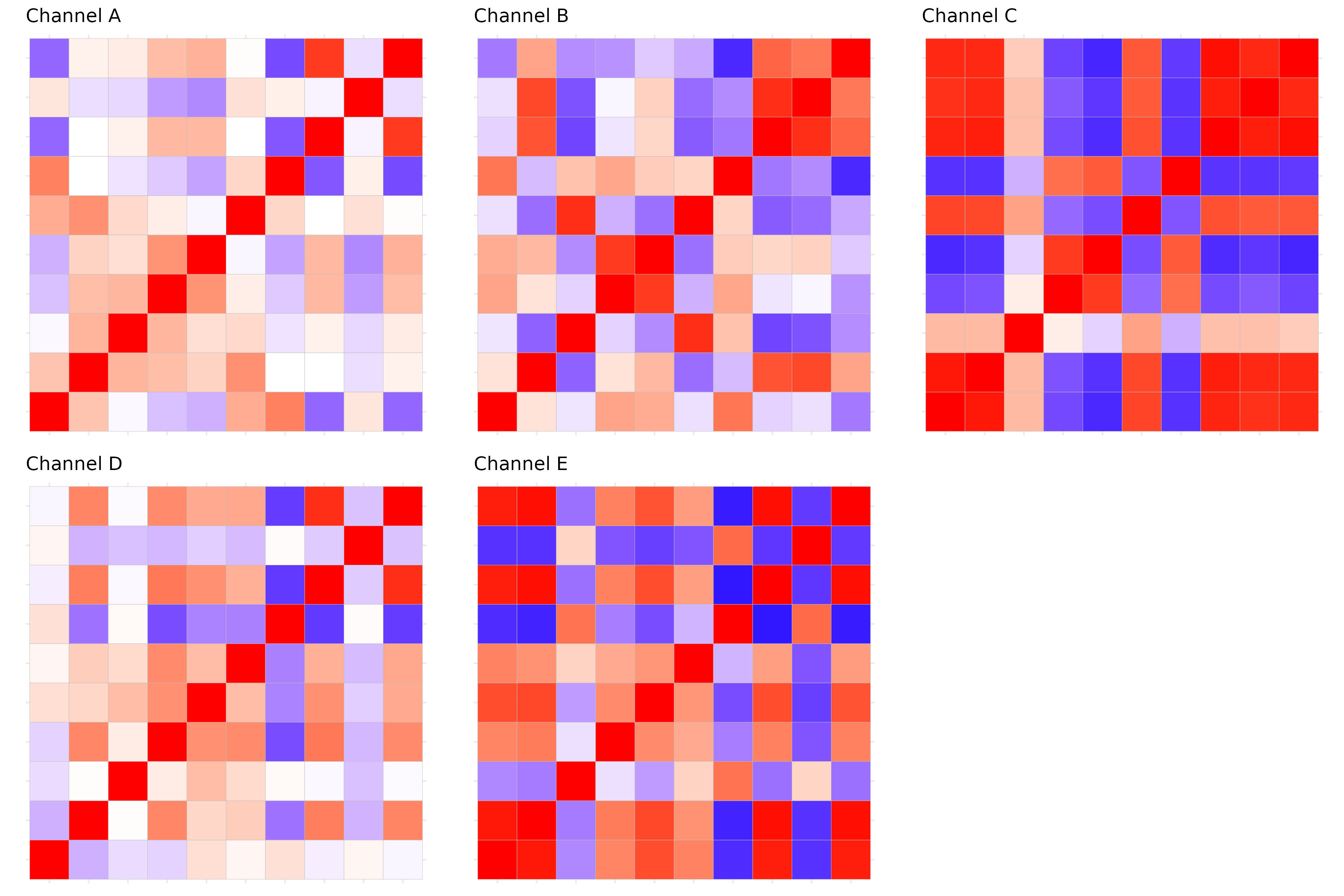}
    \end{subfigure}
 \label{fig:corr_by_ventile}
\end{figure}

Such correlation is more pronounced across some DMAs than other DMAs. Figure \ref{fig:corr_by_ventile} compares two sets of DMAs -- DMAs in the $X$th ventile of baseline sales exhibit extremely high correlation for four out of the five marketing channels, while DMAs in the $Y$th ventile exhibit relatively little correlation for all channels.\footnote{Throughout this paper, DMA IDs and channel names have been anonymized, while channel impressions have been indexed.} 

\section{Hierarchical Clustering as a Novel Solution to Multicollinearity}
As overviewed in Section \ref{Intro} and illustrated in Section \ref{Data}, multicollinearity poses a fundamental challenge in separately measuring the impact of different marketing channels. While common solutions such as shrinkage estimators, principal component regressions, and partial linear regression are helpful in prediction problems, a crucial limitation hinders their applicability to causal inference problems — they cannot provide the original causal relationships for business interpretability. 

To overcome this limitation, we propose a novel and intuitive approach that defines distances and hierarchically clusters geographic areas in a way that effectively mitigates cross-channel multicollinearity. We first calculate a pairwise distance or dissimilarity metric to summarize marketing spend correlation between geos and then use that metric to cluster the geos with moderate to strong correlation. For each channel $k$, we calculate the distance between two DMAs $i$ and $j$ as follows, where $X_{ik}$ denotes the time series of residual impressions, after eliminating the common trend and seasonality, for channel $k$ in DMA $i$. 
\[Distance_{ijk} = 1 - Correlation(X_{ik}, X_{jk})\]  
We then calculate an overall distance across all channels, which is the square-root of the sum of squared distances across the channels. 
\[Distance_{ij} = \sqrt{\sum_{k=1}^K Distance_{ijk}^2}\]

This distance measure reflects correlation between DMAs across multiple channels and is used to hierarchically cluster DMAs. We adopt a complete-linkage hierarchical clustering algorithm which works as follows:\cite{murtagh2012algorithms}\cite{reddy2018survey}
\begin{enumerate}
    \item Start with assigning each DMA to its own cluster;
    \item Then proceed iteratively, joining the two most similar clusters at each step, continuing until there is just a single cluster. Distance or dissimilarity between two clusters is based on the farthest pair.  
\end{enumerate} 

\begin{figure}[h]
  \centering
  \caption{Dendrogram Illustrating Hierarchical Clustering of DMAs}
  \includegraphics[width=\linewidth]{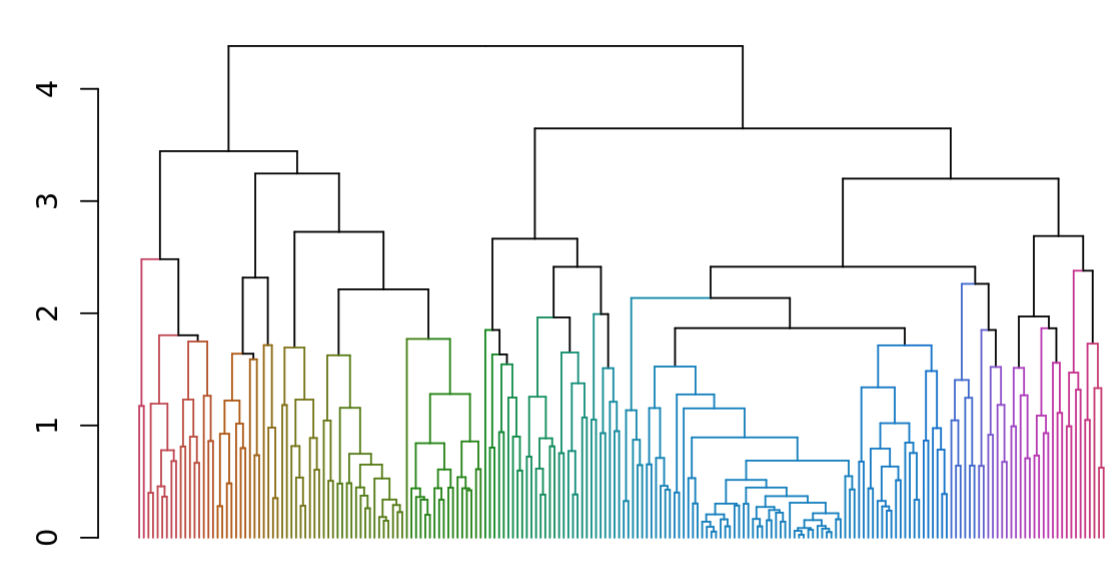}
  \label{fig:dendrogram}
\end{figure}

This algorithm produces a dendrogram in Figure \ref{fig:dendrogram}, which illustrates how DMAs are clustered at each step. The horizontal axis lays out the DMAs while the vertical axis shows the distance. This algorithm offers a lot of flexibility in how aggressively we want to cluster DMAs or how many clusters we want to have -- we can pick any cutoff distance between 0 (i.e., each DMA in a separate cluster) and 4 (i.e., all DMAs in one cluster).  

 We use a cutoff distance of 1.5, which corresponds to correlation of at least 0.33 on average for each channel and produces 42 clusters, but also consider alternative clustering strategies for sensitivity. Intuitively speaking, we group DMAs that feature moderate to strong correlation into the same cluster.

\section{Results}

\subsection{Hierarchical Clustering of DMAs}
\begin{figure}
  \centering
  \caption{Heat Maps of Channel Residual Impressions Across DMAs Over Time}
    \begin{subfigure}[b]{0.5\textwidth}
        \centering
        \caption{DMAs in Cluster 1}
        \includegraphics[width=\linewidth]{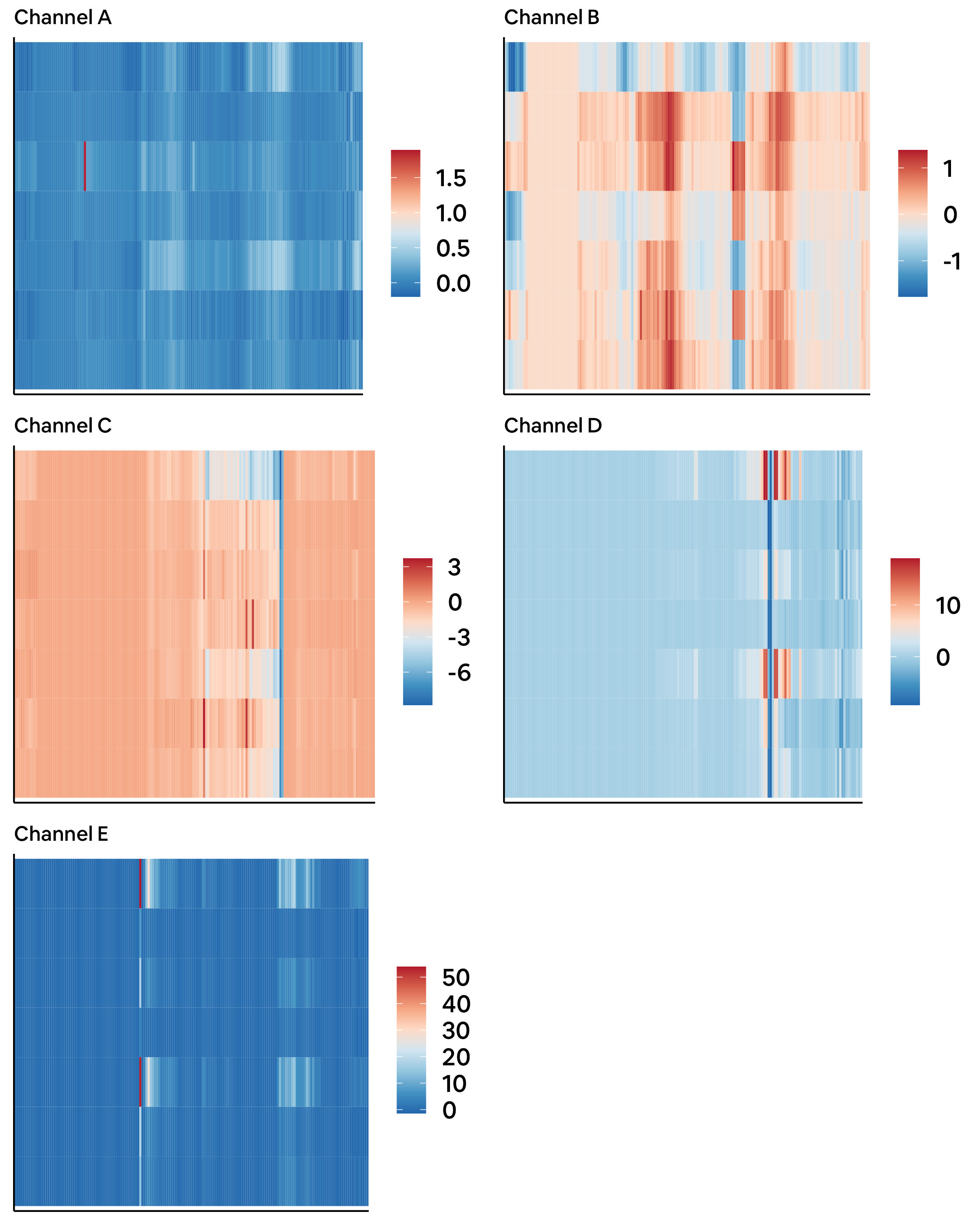}   
    \end{subfigure}
    \hfill
    \begin{subfigure}[b]{0.5\textwidth}
        \centering
        \caption{DMAs in Cluster 4}
        \includegraphics[width=\linewidth]{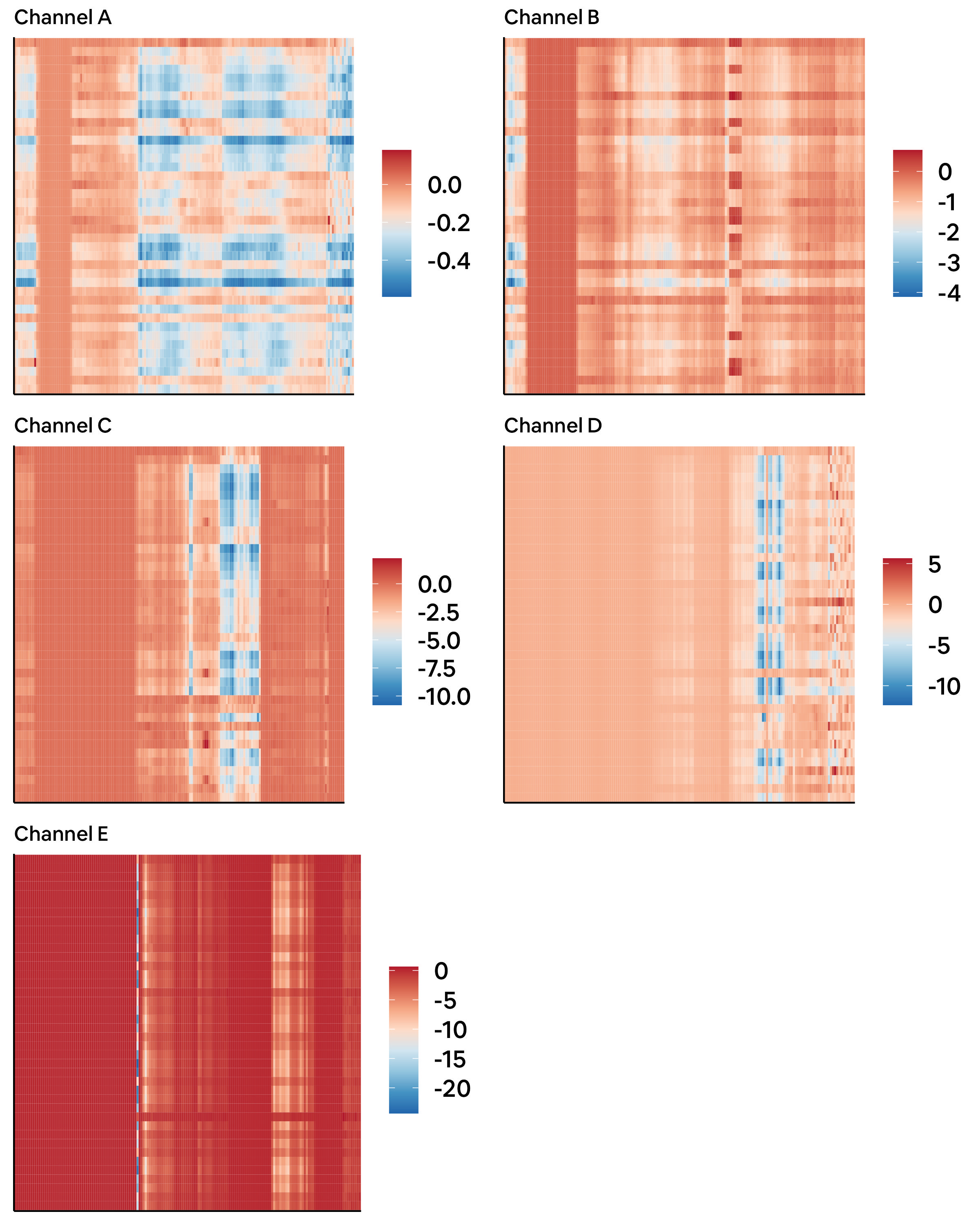} 
    \end{subfigure}
 \label{fig:heatmap_by_cluster}
\end{figure}

The hierarchical clustering approach produces intuitive results. We visualize channel impressions over time across DMAs within each cluster, confirming that DMAs within the same cluster tend to have highly correlated impressions over time for at least some channels. Figure \ref{fig:heatmap_by_cluster} exemplifies such patterns using one small cluster (Cluster 1) and one larger cluster (Cluster 4). Each chart illustrates the variation in channel residual impression over time (the horizontal axis) and across DMAs (the vertical axis). Within each cluster, the color patterns over time are quite similar across DMAs for most channels, reflecting moderate to high correlation. 
 
\subsection{Descriptive Evidence Shows Clustering Mitigates Multicollinearity}
\begin{figure}[h]
  \centering
  \caption{Clustering Reduces Cross-Channel Correlation}
  \includegraphics[width=\linewidth]{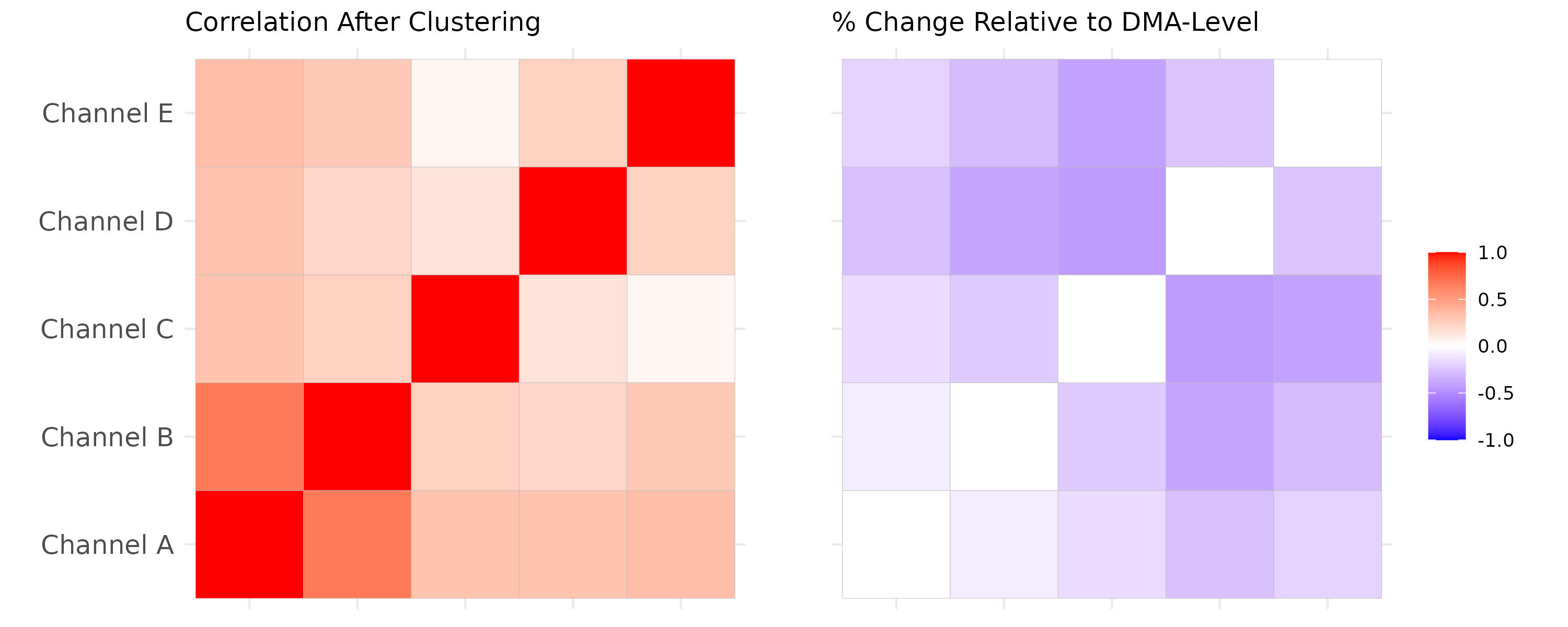}
  \label{fig:corr_change}
\end{figure}

Descriptive evidence confirms our intuition that hierarchical clustering can effectively mitigate collinearity. By grouping moderately to highly correlated DMAs into the same cluster, we have significantly reduced correlation in residual channel impressions. As Figure \ref{fig:corr_change} visualizes, correlation decreased generally across channels, by $8\%$ to $43\%$.  

\begin{figure}[h]
  \centering
  \caption{Variation in Residual Channel Impressions Across Clusters Over Time}
  \includegraphics[width=\linewidth]{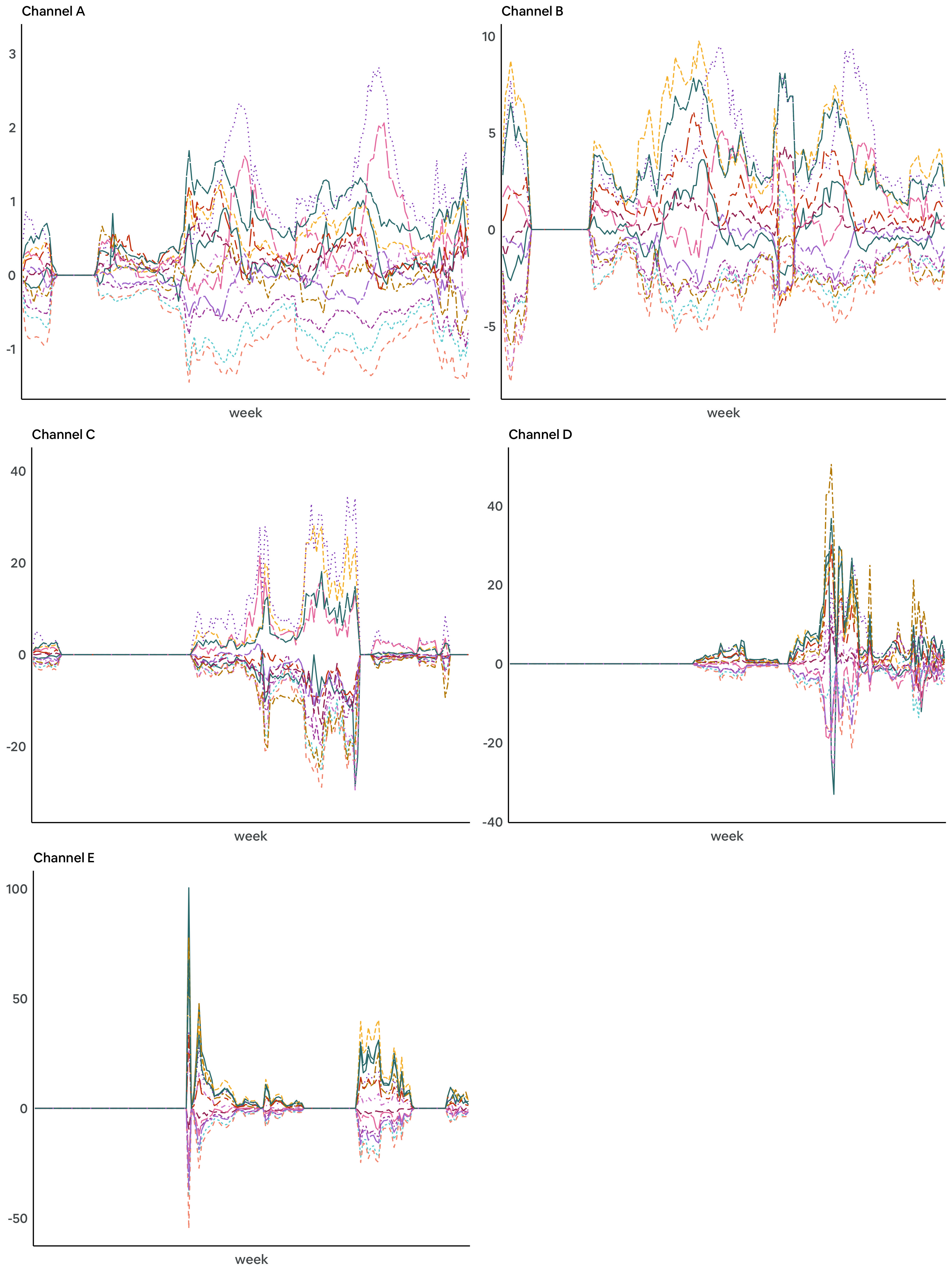}
  \label{fig:variation_by_cluster_over_time}
\end{figure}

Further, as Figure \ref{fig:variation_by_cluster_over_time} demonstrates, clustering preserves variation in channel impressions both (A) within clusters over time and (B) across clusters within the same time period. This is promising for separately identifying the impact of different channels using panel data at the cluster-week level. When testing alternative clustering strategies, we consider the reduction in correlation and the preservation of variation as two important criteria. 

\subsection{Regression Results Confirm Clustering Alleviates Multicollinearity}
Panel linear regression analysis affirms the effectiveness of this hierarchical clustering method in mitigating collinearity and facilitating the separate identification of the impact of different marketing channels. After clustering, channel coefficient estimates are no longer subject to the problem of flipped signs when included together with other channels, and instead produce mostly intuitive results. 

Table \ref{table:plm} summarizes panel linear regression results before and after clustering, with geo (DMA or cluster) fixed effects and week fixed effects included throughout the different specifications.\footnote{All coefficient estimates have been scaled by a constant.} As Column 1 summarizes, most channel coefficients are negative if we use DMA-level data, while they would be positive if included individually. After clustering, in Column 2, the results become mostly intuitive -- all four lower-funnel channels have positive estimated impact on sales (three of which are significant at 0.001 level). The remaining channel with a negative coefficient is upper-funnel, where we expect extreme difficulty in detecting a lower funnel impact. Finally, these findings are robust to weighting the cluster based on the natural log of their baseline size. 

\begin{table} 
\caption{Panel Linear Regression Summary}
    \label{table:plm}
\centering
\begin{tabular}[t]{lccc}
\toprule
  & DMA Level & Cluster Level & Cluster Level, Weighted\\
\midrule
channel\_a & \num{5.570}*** & \num{5.041}*** & \num{5.227}***\\
 & (\num{0.075}) & (\num{0.149}) & (\num{0.146})\\
channel\_b & \num{-0.055}*** & \num{0.128}*** & \num{0.097}***\\
 & (\num{0.014}) & (\num{0.028}) & (\num{0.027})\\
channel\_c & \num{-0.003} & \num{0.043}*** & \num{0.038}***\\
 & (\num{0.004}) & (\num{0.008}) & (\num{0.007})\\
channel\_d & \num{0.007} & \num{0.007} & \num{0.009}\\
 & (\num{0.005}) & (\num{0.010}) & (\num{0.010})\\
channel\_e & \num{-0.011}*** & \num{-0.023}** & \num{-0.025}***\\
 & (\num{0.003}) & (\num{0.008}) & (\num{0.007})\\
\midrule
Num.Obs. & \num{35700} & \num{7140} & \num{7140}\\
R2 & \num{0.862} & \num{0.946} & \num{0.946}\\
R2 Adj. & \num{0.861} & \num{0.944} & \num{0.944}\\
AIC & \num{113008.3} & \num{34358.7} & \num{48912.9}\\
BIC & \num{114492.8} & \num{35561.6} & \num{50115.8}\\
RMSE & \num{1.17} & \num{2.62} & \num{7.27}\\
\bottomrule
\multicolumn{4}{l}{\rule{0pt}{1em}+ p $<$ 0.1, * p $<$ 0.05, ** p $<$ 0.01, *** p $<$ 0.001}\\
\end{tabular}
\end{table}

\subsection{Bayesian Model Results Using Cluster-Level Data}
\begin{figure}[h]
  \centering
  \caption{Posterior Distributions of Parameters}
  \includegraphics[width=\linewidth]{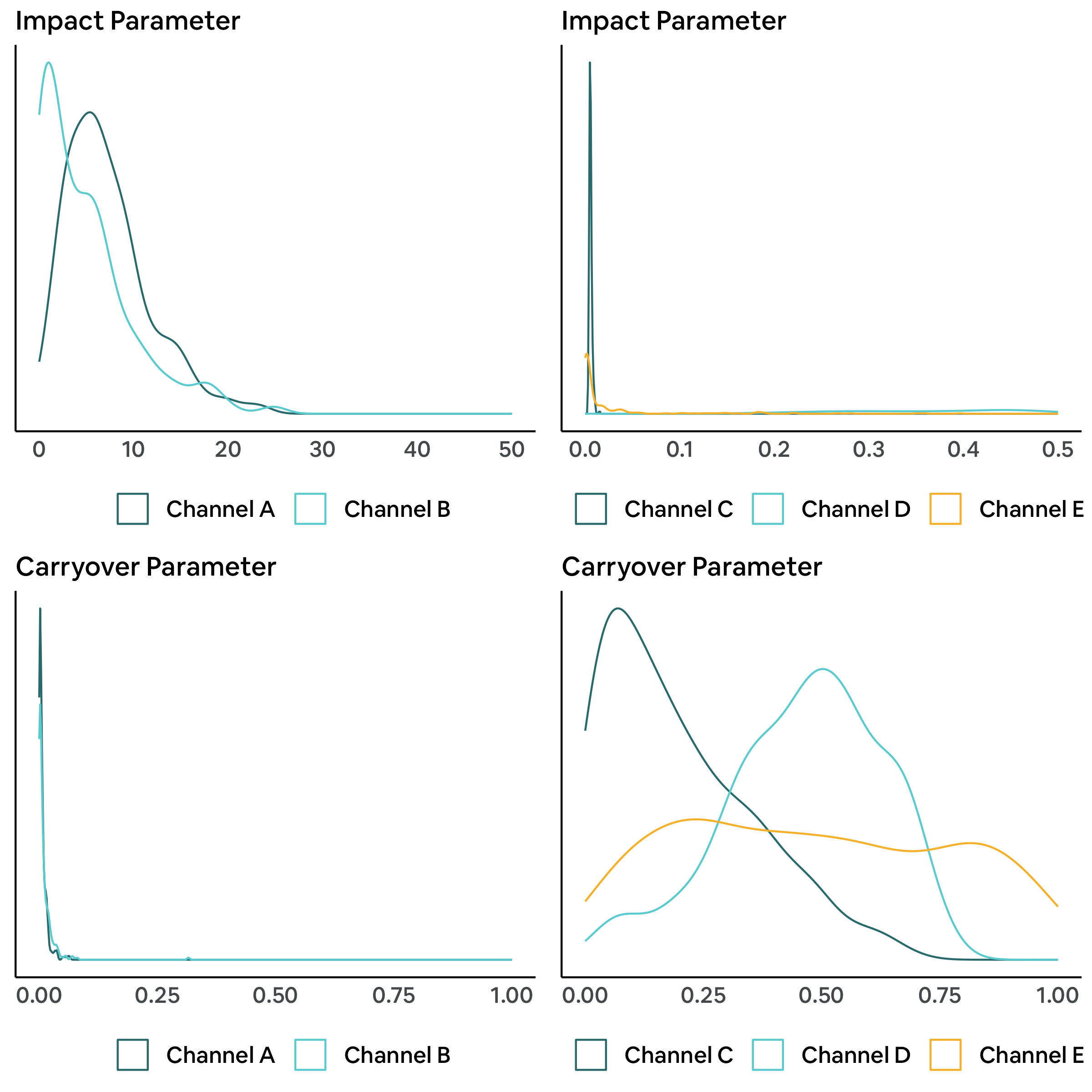}
  \label{fig:bayesian_estimates}
\end{figure}

Now that both descriptive evidence and frequentist regression analysis affirm our hierarchical clustering approach effectively mitigates collinearity, we move forward to estimate the Bayesian model described in Section \ref{BayesianModel} using data at the cluster level. Similar to the frequentist regressions, the Bayesian model also produces intuitive results, even with uninformative priors. Figure \ref{fig:bayesian_estimates} visualizes the posterior distribution of channel-specific impact parameters ($\beta$) and carryover rate parameters ($\tau$). Consistently with frequentist regression results earlier, we estimate a higher impact for channels A and B than the other channels.\footnote{Note that the impact parameter estimates here should be interpreted differently from the frequentist panel linear regressions above, because the Bayesian structural model also estimates parameters that transforms the impressions for each channel into adstock based on the lag, carryover, and shape parameters. But we can still make broad comparisons of the impact parameter across channels, taking into account the other parameters.} Furthermore, the carryover estimates are also intuitive and consistent with our previous learning and knowledge about the different channels. For example, we expect some carryover for Channels C, D, and E, but not for Channels A and B, and it is reassuring that the estimates confirm that understanding even though we are using uninformative priors.\footnote{It is expected that the variance is high for the estimates of Channel E, such as the carryover parameter. Unlike the other channels, Channel E is upper funnel, so it is especially difficult to estimate its impact on lower funnel conversions.}   

\section{Discussion of Broader Applications}
We have focused on an application in marketing mix modeling to demonstrate how and why hierarchical clustering  can mitigate multicollinearity. However, this method is not constrained to the setting of marketing and instead is generally applicable to observational causal inference problems featuring multicollinearity. In our example, marketing data properties motivated us to cluster geographic units based on correlation in marketing activities. In other settings, one can decide which dimensions and criteria to use for clustering based on relevant data properties. The dimension to cluster data does not always have to be geographic. Futher more, in some settings, natural clusters might exist for one to consider. 

For example, in the context of Customer Service, we would like to understand how customers' each interaction with our support agents contribute to the long term retention. However, oftentimes, these interaction experience metrics are highly correlated, such as wait time and abandon rate, etc. In this case, we can leverage clustering to segment customer issue types into groups that have different degrees of correlation between wait and abandon rate. 

\section{Conclusion}
In this paper, we propose to employ hierarchical clustering as an innovative and effective approach to address multicollinearity in regressional causal inference studies. It has several advantages. Firstly, hierarchical clustering provides a systematic and comprehensive method for identifying clusters that exhibit varying levels of multicollinearity, thus reducing the correlation of covariates across clusters. Furthermore, clustering circumvents the need to transform data into non-interpretable entities, as required by techniques such as Principal Component Analysis or Partial Linear Regressions. This ensures that the interpretability and meaningfulness of the variables are preserved throughout the analysis. In addition to its effectiveness, the proposed methodology is characterized by its ease of implementation. It can be readily applied to diverse applications facing similar challenges related to multicollinearity. The key lies in understanding the inherent properties of the data to define an appropriate distance metric for clustering that effectively reduces multicollinearity. This research contributes to enhancing the robustness and reliability of regressional causal inference studies. 

\section{Acknowledgement}
The authors would like to thank Carolina Barcenas (Airbnb, Inc.), Mike Anderson (Google LLC), Fei Cen (Google LLC) for their help and guidance on this project.

\nocite{chan2017challenges}
\nocite{wang2017hierarchical}
\nocite{blake2015consumer}
\nocite{hardle2000partially}
\nocite{chen2021hierarchical}
\nocite{daoud2017multicollinearity}
\nocite{du2019causally}
\nocite{imbens_rubin_2015}
\nocite{hoerl1970ridge}
\nocite{narayanan2014position}
\nocite{vaver2017introduction}
\nocite{berman2018beyond}
\nocite{wolfe2011marketing}
\nocite{thomas2020spillovers}


\bibliography{sample-base}


\begin{thebibliography}{18}


\ifx \showCODEN    \undefined \def \showCODEN     #1{\unskip}     \fi
\ifx \showDOI      \undefined \def \showDOI       #1{#1}\fi
\ifx \showISBNx    \undefined \def \showISBNx     #1{\unskip}     \fi
\ifx \showISBNxiii \undefined \def \showISBNxiii  #1{\unskip}     \fi
\ifx \showISSN     \undefined \def \showISSN      #1{\unskip}     \fi
\ifx \showLCCN     \undefined \def \showLCCN      #1{\unskip}     \fi
\ifx \shownote     \undefined \def \shownote      #1{#1}          \fi
\ifx \showarticletitle \undefined \def \showarticletitle #1{#1}   \fi
\ifx \showURL      \undefined \def \showURL       {\relax}        \fi
\providecommand\bibfield[2]{#2}
\providecommand\bibinfo[2]{#2}
\providecommand\natexlab[1]{#1}
\providecommand\showeprint[2][]{arXiv:#2}

\bibitem[Berman(2018)]%
        {berman2018beyond}
\bibfield{author}{\bibinfo{person}{Ron Berman}.}
  \bibinfo{year}{2018}\natexlab{}.
\newblock \showarticletitle{Beyond the last touch: Attribution in online
  advertising}.
\newblock \bibinfo{journal}{\emph{Marketing Science}} \bibinfo{volume}{37},
  \bibinfo{number}{5} (\bibinfo{year}{2018}), \bibinfo{pages}{771--792}.
\newblock


\bibitem[Blake et~al\mbox{.}(2015)]%
        {blake2015consumer}
\bibfield{author}{\bibinfo{person}{Thomas Blake}, \bibinfo{person}{Chris
  Nosko}, {and} \bibinfo{person}{Steven Tadelis}.}
  \bibinfo{year}{2015}\natexlab{}.
\newblock \showarticletitle{Consumer heterogeneity and paid search
  effectiveness: A large-scale field experiment}.
\newblock \bibinfo{journal}{\emph{Econometrica}} \bibinfo{volume}{83},
  \bibinfo{number}{1} (\bibinfo{year}{2015}), \bibinfo{pages}{155--174}.
\newblock


\bibitem[Chan and Perry(2017)]%
        {chan2017challenges}
\bibfield{author}{\bibinfo{person}{David Chan} {and} \bibinfo{person}{Mike
  Perry}.} \bibinfo{year}{2017}\natexlab{}.
\newblock \showarticletitle{Challenges and opportunities in media mix
  modeling}.
\newblock  (\bibinfo{year}{2017}).
\newblock


\bibitem[Chen et~al\mbox{.}(2021)]%
        {chen2021hierarchical}
\bibfield{author}{\bibinfo{person}{Hao Chen}, \bibinfo{person}{Minguang Zhang},
  \bibinfo{person}{Lanshan Han}, {and} \bibinfo{person}{Alvin Lim}.}
  \bibinfo{year}{2021}\natexlab{}.
\newblock \showarticletitle{Hierarchical marketing mix models with sign
  constraints}.
\newblock \bibinfo{journal}{\emph{Journal of Applied Statistics}}
  \bibinfo{volume}{48}, \bibinfo{number}{13-15} (\bibinfo{year}{2021}),
  \bibinfo{pages}{2944--2960}.
\newblock


\bibitem[Daoud(2017)]%
        {daoud2017multicollinearity}
\bibfield{author}{\bibinfo{person}{Jamal~I Daoud}.}
  \bibinfo{year}{2017}\natexlab{}.
\newblock \showarticletitle{Multicollinearity and regression analysis}. In
  \bibinfo{booktitle}{\emph{Journal of Physics: Conference Series}},
  Vol.~\bibinfo{volume}{949}. IOP Publishing, \bibinfo{pages}{012009}.
\newblock


\bibitem[Du et~al\mbox{.}(2019)]%
        {du2019causally}
\bibfield{author}{\bibinfo{person}{Ruihuan Du}, \bibinfo{person}{Yu Zhong},
  \bibinfo{person}{Harikesh Nair}, \bibinfo{person}{Bo Cui}, {and}
  \bibinfo{person}{Ruyang Shou}.} \bibinfo{year}{2019}\natexlab{}.
\newblock \showarticletitle{Causally driven incremental multi touch attribution
  using a recurrent neural network}.
\newblock \bibinfo{journal}{\emph{arXiv preprint arXiv:1902.00215}}
  (\bibinfo{year}{2019}).
\newblock


\bibitem[H{\"a}rdle et~al\mbox{.}(2000)]%
        {hardle2000partially}
\bibfield{author}{\bibinfo{person}{Wolfgang H{\"a}rdle}, \bibinfo{person}{Hua
  Liang}, {and} \bibinfo{person}{Jiti Gao}.} \bibinfo{year}{2000}\natexlab{}.
\newblock \bibinfo{booktitle}{\emph{Partially linear models}}.
\newblock \bibinfo{publisher}{Springer Science \& Business Media}.
\newblock


\bibitem[Hoerl and Kennard(1970)]%
        {hoerl1970ridge}
\bibfield{author}{\bibinfo{person}{Arthur~E Hoerl} {and}
  \bibinfo{person}{Robert~W Kennard}.} \bibinfo{year}{1970}\natexlab{}.
\newblock \showarticletitle{Ridge regression: applications to nonorthogonal
  problems}.
\newblock \bibinfo{journal}{\emph{Technometrics}} \bibinfo{volume}{12},
  \bibinfo{number}{1} (\bibinfo{year}{1970}), \bibinfo{pages}{69--82}.
\newblock


\bibitem[Imbens and Rubin(2015)]%
        {imbens_rubin_2015}
\bibfield{author}{\bibinfo{person}{Guido~W. Imbens} {and}
  \bibinfo{person}{Donald~B. Rubin}.} \bibinfo{year}{2015}\natexlab{}.
\newblock \bibinfo{booktitle}{\emph{Causal Inference for Statistics, Social,
  and Biomedical Sciences: An Introduction}}.
\newblock \bibinfo{publisher}{Cambridge University Press}.
\newblock
\urldef\tempurl%
\url{https://doi.org/10.1017/CBO9781139025751}
\showDOI{\tempurl}


\bibitem[Jin et~al\mbox{.}(2017)]%
        {jin2017bayesian}
\bibfield{author}{\bibinfo{person}{Yuxue Jin}, \bibinfo{person}{Yueqing Wang},
  \bibinfo{person}{Yunting Sun}, \bibinfo{person}{David Chan}, {and}
  \bibinfo{person}{Jim Koehler}.} \bibinfo{year}{2017}\natexlab{}.
\newblock \showarticletitle{Bayesian methods for media mix modeling with
  carryover and shape effects}.
\newblock  (\bibinfo{year}{2017}).
\newblock


\bibitem[Murtagh and Contreras(2012)]%
        {murtagh2012algorithms}
\bibfield{author}{\bibinfo{person}{Fionn Murtagh} {and} \bibinfo{person}{Pedro
  Contreras}.} \bibinfo{year}{2012}\natexlab{}.
\newblock \showarticletitle{Algorithms for hierarchical clustering: an
  overview}.
\newblock \bibinfo{journal}{\emph{Wiley Interdisciplinary Reviews: Data Mining
  and Knowledge Discovery}} \bibinfo{volume}{2}, \bibinfo{number}{1}
  (\bibinfo{year}{2012}), \bibinfo{pages}{86--97}.
\newblock


\bibitem[Narayanan and Kalyanam(2014)]%
        {narayanan2014position}
\bibfield{author}{\bibinfo{person}{Sridhar Narayanan} {and}
  \bibinfo{person}{Kirthi Kalyanam}.} \bibinfo{year}{2014}\natexlab{}.
\newblock \bibinfo{booktitle}{\emph{Position effects in search advertising: A
  regression discontinuity approach}}.
\newblock \bibinfo{type}{{T}echnical {R}eport}. \bibinfo{institution}{Working
  paper}.
\newblock


\bibitem[Ng et~al\mbox{.}(2021)]%
        {ng2021bayesian}
\bibfield{author}{\bibinfo{person}{Edwin Ng}, \bibinfo{person}{Zhishi Wang},
  {and} \bibinfo{person}{Athena Dai}.} \bibinfo{year}{2021}\natexlab{}.
\newblock \showarticletitle{Bayesian Time Varying Coefficient Model with
  Applications to Marketing Mix Modeling}.
\newblock \bibinfo{journal}{\emph{arXiv preprint arXiv:2106.03322}}
  (\bibinfo{year}{2021}).
\newblock


\bibitem[Reddy and Vinzamuri(2018)]%
        {reddy2018survey}
\bibfield{author}{\bibinfo{person}{Chandan~K Reddy} {and}
  \bibinfo{person}{Bhanukiran Vinzamuri}.} \bibinfo{year}{2018}\natexlab{}.
\newblock \showarticletitle{A survey of partitional and hierarchical clustering
  algorithms}.
\newblock In \bibinfo{booktitle}{\emph{Data clustering}}.
  \bibinfo{publisher}{Chapman and Hall/CRC}, \bibinfo{pages}{87--110}.
\newblock


\bibitem[Thomas(2020)]%
        {thomas2020spillovers}
\bibfield{author}{\bibinfo{person}{Michael Thomas}.}
  \bibinfo{year}{2020}\natexlab{}.
\newblock \showarticletitle{Spillovers from mass advertising: An identification
  strategy}.
\newblock \bibinfo{journal}{\emph{Marketing Science}} \bibinfo{volume}{39},
  \bibinfo{number}{4} (\bibinfo{year}{2020}), \bibinfo{pages}{807--826}.
\newblock


\bibitem[Vaver and Zhang(2017)]%
        {vaver2017introduction}
\bibfield{author}{\bibinfo{person}{Jon Vaver} {and} \bibinfo{person}{Stephanie
  Shin-Hui Zhang}.} \bibinfo{year}{2017}\natexlab{}.
\newblock \showarticletitle{Introduction to the Aggregate Marketing System
  Simulator}.
\newblock  (\bibinfo{year}{2017}).
\newblock


\bibitem[Wang et~al\mbox{.}(2017)]%
        {wang2017hierarchical}
\bibfield{author}{\bibinfo{person}{Yueqing Wang}, \bibinfo{person}{Yuxue Jin},
  \bibinfo{person}{Yunting Sun}, \bibinfo{person}{David Chan}, {and}
  \bibinfo{person}{Jim Koehler}.} \bibinfo{year}{2017}\natexlab{}.
\newblock \showarticletitle{A hierarchical Bayesian approach to improve media
  mix models using category data}.
\newblock  (\bibinfo{year}{2017}).
\newblock


\bibitem[Wolfe~Sr and Crotts(2011)]%
        {wolfe2011marketing}
\bibfield{author}{\bibinfo{person}{Michael~J Wolfe~Sr} {and}
  \bibinfo{person}{John~C Crotts}.} \bibinfo{year}{2011}\natexlab{}.
\newblock \showarticletitle{Marketing mix modeling for the tourism industry: A
  best practices approach}.
\newblock \bibinfo{journal}{\emph{International Journal of Tourism Sciences}}
  \bibinfo{volume}{11}, \bibinfo{number}{1} (\bibinfo{year}{2011}),
  \bibinfo{pages}{1--15}.
\newblock


\end{thebibliography}


\end{document}